\documentstyle[11pt,aspconf]{article}

\begin{document}

\title{HUT Observations of the Lyman Limit in AGN:
	Implications for Unified Models}

\author{G.\ Kriss, J. Krolik, J. Grimes, Z. Tsvetanov, B. Espey, W. Zheng,\\
and A. Davidsen}
\affil{Department of Physics and Astronomy, The Johns Hopkins University,
    Baltimore, MD 21218}

\begin{abstract}
We observed a total of 13 low-redshift AGN with the Hopkins Ultraviolet
Telescope (HUT) during the course of the Astro-1 and Astro-2 missions.
Of these, 4 show intrinsic Lyman limit absorption---NGC 1068,
NGC~4151, NGC~3516 and NGC~3227.  All galaxies with optically thick Lyman
limits
have extended narrow-line regions with bi-polar morphologies.
All AGN with no absorption are Seyfert 1's with compact NLR's.
These observations support geometrical shadowing as the means for collimating
the ionizing radiation in unified models of AGN, most likely in a photoionized
atmosphere above the obscuring torus.
\end{abstract}


\section{Introduction}

An advantage of being one of the last speakers is that my preceding colleagues
have given you an excellent introduction.
So, I need only remind you of a few key points.
The most important feature to remember from Wei Zheng's description of
HUT is its sensitivity below 1200 \AA.
This allows us to see the Lyman limit region in low redshift AGN.
A glimpse of the far-ultraviolet properties of these bright, nearby AGN
is important because our whole concept of a unified model is based on
low-redshift objects.  Nearby AGN are bright, allowing spectropolarimetry
and high-resolution X-ray spectroscopy, and the good angular scale allows us to
resolve the narrow-line region (NLR) with HST and from the ground.

Unified models of AGN use a combination of
obscuration, reflection, and orientation to explain the different appearances
of Seyfert 1 and Seyfert 2 galaxies (see the review by Antonucci 1993).
This picture implies some anisotropy for the escaping ionizing radiation.
As Andrew Wilson showed you yesterday, this has important consequences for
the morphology of the NLR.  If NLR gas is spherically distributed about the
nuclear region, the escaping radiation will produce biconical structures
projected onto the sky in obscured AGN (Type 2) and more compact, more
symmetric structures in those viewed pole on (Type 1's).
By and large, this is what is observed (e.g., Pogge 1989; Evans et al. 1994).

In more general terms,
one could classify the means of collimating the ionizing radiation in
one of two categories: intrinsic anisotropy, or geometrical shadowing.
Examples of intrinsically anisotropic radiation would be that due to
relativistic beaming or the radiation from an accretion funnel in a
geometrically thick accretion disk.
Geometrical shadows that collimate the radiation can be produced not only by
the obscuring torus, but also by the wind from an accretion disk or by a
toroidal configuration of broad-line region clouds.
{\it All examples of shadowing predict we should see optically thick neutral
hydrogen absorption at the Lyman limit.}

\begin{figure}
\plotfiddle{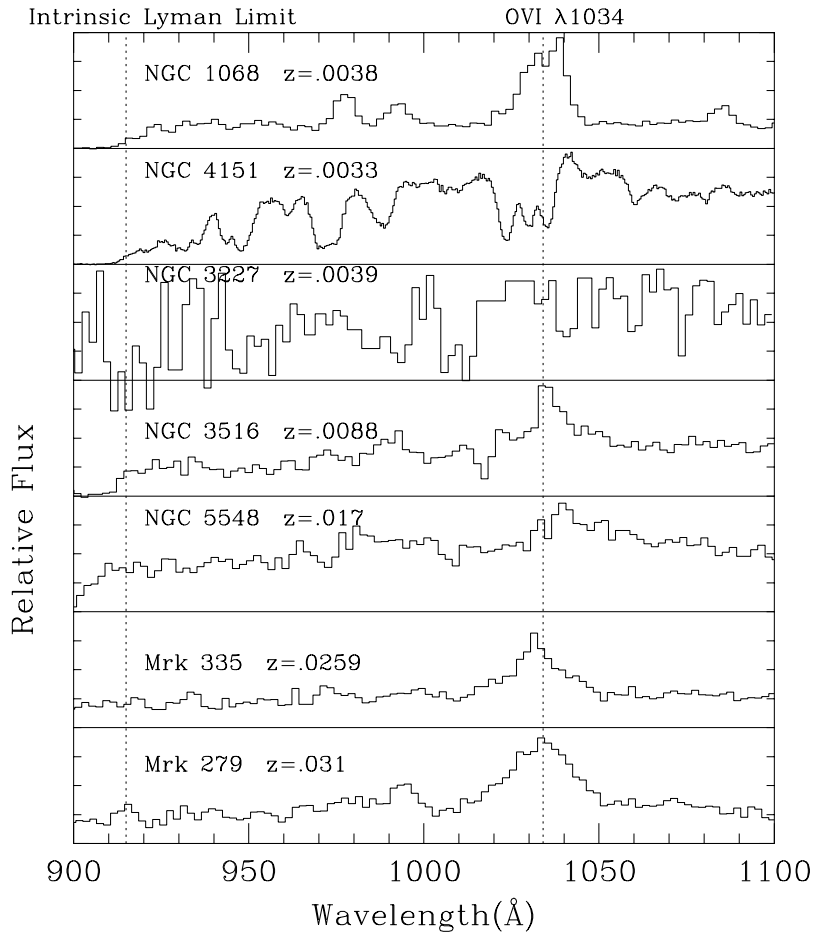}{4.63in}{0.0}{58}{58}{-187}{-56}
\caption{Representative HUT spectra of the Lyman limit region in
low-redshift AGN.  All spectra are shown in the object's rest frame.}
\end{figure}

\section{HUT Observations}

Of 13 AGN with $z < 0.3$ observed with HUT, only 4 show intrinsic Lyman limits.
Table 1 summarizes the HUT Astro-1 and Astro-2 observations, and
Figure 1 shows representative spectra.
All objects with Lyman limit absorption
(NGC~1068, NGC~4151, NGC~3516, and NGC~3227) have
narrow-line regions with bipolar morphologies.
The remaining 9 AGN are Seyfert 1's or low-redshift quasars with
point-like NLR's as imaged with HST or from the ground.
{\it Seeing absorption only in objects with bipolar NLR's favors shadowing for
collimation over an intrinsically beamed mechanism.}

\begin{deluxetable}{l c c c c}
\tablecaption{AGN with HUT Lyman Limit Observations. \label{tbl-1}}
\tablehead{
\colhead{Object} & \colhead{Type} & \colhead{$z$} & \colhead{NLR Morphology} &
\colhead{Lyman Limit?}}
\startdata
NGC 1068 & 2 & 0.0038 & bi-cone & yes \nl
NGC 4151 & 1 & 0.0033 & bi-cone & yes \nl
NGC 3227 & 1 & 0.0039 & bi-cone & yes (95\% confidence) \nl
NGC 3516 & 1 & 0.0088 & S-shape & yes \nl
NGC 5548 & 1 & 0.017\phantom{0} & point-like & no \nl
Mrk 335  & 1 & 0.026\phantom{0} & point-like & no \nl
Mrk 279  & 1 & 0.031\phantom{0} & point-like & no \nl
Mrk 509  & 1 & 0.035\phantom{0} & point-like & no \nl
Mrk 478  & 1 & 0.077\phantom{0} & point-like & no \nl
PG1211+143 & 1 & 0.081\phantom{0} & ? & no \nl
PG1351+640 & 1 & 0.088\phantom{0} & ? & no \nl
3C 273     & 1 & 0.158\phantom{0} & jet & no \nl
E1821+64   & 1 & 0.297\phantom{0} & point-like & no \nl
\enddata
\end{deluxetable}

\section{The Location of the Collimator}

The next question is, if obscuration collimates the light, can we pinpoint
the location of the absorbing gas?  Monitoring variability in absorption
features allows one to measure two timescales---the ionization timescale and
the
recombination timescale---each of which allows some measure of the distance
of the absorbing gas from the central source.
To apply this technique, we observed NGC~4151 with HUT 6 times at intervals
of 2--3 days during the Astro-2 mission, and we got lucky.
Our data (Fig. 2) show an approximate $\delta$-function blip in the continuum
flux and the change of the neutral hydrogen column in response.
When the continuum flux rises (a 60\% increase,
or $\Delta L = 1.4 \times 10^{43}~\rm erg~s^{-1}$),
we see a nearly instantaneous drop in the
neutral hydrogen column ($\Delta N_{HI} = 7.5 \times 10^{17}~\rm cm^{-2}$)
giving an ionization timescale of $< 2$ days, the limit of our sampling.
After the continuum returns to its mean intensity, the neutral hydrogen column
gradually increases with an e-folding timescale of $\sim 10$ days.

\begin{figure}
\plotfiddle{"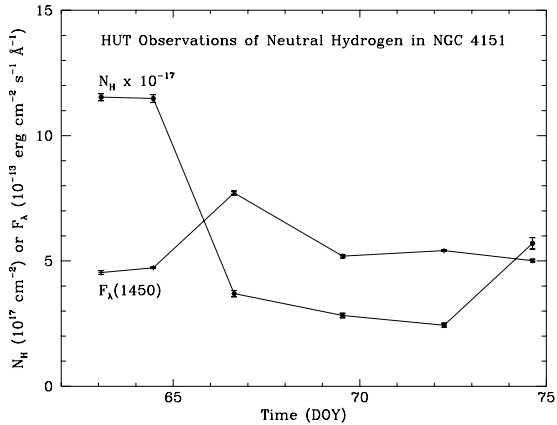"}
{1.9in}{-90}{30}{30}{-115}{165}
\caption{ The variation in the neutral hydrogen column
from HUT observations of NGC 4151 during Astro-2 is shown along with
the continuum intensity at 1450 \AA.
The x-axis is in day of the year 1995. Error bars, barely visible around
each data point, are 1 $\sigma$.}
\end{figure}

The decrease in the neutral hydrogen column, $\Delta N_{HI}$,
and the time, $\Delta t$, required to respond to the continuum increase,
$\Delta L$, provide a direct measure of the number of ionizing photons
and indirectly give a measure of the distance of the absorber from the source.
Since $\Delta N_{HI}~=~( {\rm \#~ionizing~photons}) / (4 \pi r^2)$, and
$\Delta L = ( {\rm \#~ionizing~photons}) \times h\bar{\nu} / \Delta t$,
the distance of the absorber from the source can be found as
$r = [ (\Delta L \Delta t) / (4 \pi \Delta N_{HI} h\bar{\nu}) ]^{1/2}$.
For our observations of NGC~4151, an upper limit of 2 days on the ionization
timescale translates into an upper limit of 30 pc on the radius of the
absorbing gas.

The recombination timescale,
$t_{rec} = (1 / n_e \alpha_{rec}) (n_{HI} / n_{HII} )$,
is often cited as a measure of the density of
absorbing material, but this is true only if one knows the ionization state
of the gas (Krolik \& Kriss 1995).
The neutral fraction $n_{HI} / n_{HII}$ depends on
the ionization parameter $U = L / (4 \pi r^2 h \bar{\nu} (n_e + n_p))$ as
$n_{HI} / n_{HII} = Q / U$,
where Q is typically $\sim 4.5 \times 10^{-6}$,
determined by atomic data and the shape of the ionizing spectrum.
These relations can be used to obtain a distance
estimate for the gas based on the observed recombination time:
$$ r = 62~{\rm pc}~\left( \frac{\Delta t}{10~\rm d} \right)^{1/2} \left(
\frac{L}{2.3 \times 10^{43}~\rm erg~s^{-1}} \right)^{1/2} \left(
\frac{15,000~\rm K}{T} \right)^{0.4} \left( \frac{4.5 \times 10^{-6}}{Q}
\right)^{1/2} .$$
For our observations of NGC~4151, the absorbing gas is located far from the
central source in the NLR.  The absorbing gas may be outflowing material
driven off the surface of the obscuring torus.

Given the limited sampling and the simplified treatment shown here, I would
view these numbers as only order of magnitude estimates.
They show, however, that the the absorbing gas is likely to be far from the
central source, at or beyond the location of the obscuring torus.
Our observations illustrate the power of monitoring absorption variability,
and this technique offers real possibilities for learning more about
both UV and X-ray absorbing gas.

\acknowledgments

This work was supported by NASA Contract NAS 5-27000 and NASA LTSA grant
NAGW-4443 to the Johns Hopkins University.

\end{document}